\documentclass[prc,aps,preprint,showpacs,nofootinbib,showkeys,eqsecnum,tightenlines]{revtex4}    
\usepackage{epsfig}
\def\be{\begin{equation}}
\def\ee{\end{equation}}
\def\bea{\begin{eqnarray}}
\def\eea{\end{eqnarray}}

\def\P{\vec P}

\def\s2q2{(\vec \sigma_2 \cdot \vec q_2)}

\newcommand{\noi}{\noindent}

\def\Journal#1#2#3#4{{#1}{\bf #2} (#4) #3 }
\def\NPA{{ Nucl. Phys.\,\,} \bf A}
\def\NPB{{ Nucl. Phys.\,\,} \bf B}
\def\NP{{ Nucl. Phys.\,\,}}
\def\PRL{ Phys. Rev. Lett.\,\,}
\def\PRC{{ Phys. Rev.\,\,}  C}

\def\PRW{ Phys. Rev.\,\,}
\def\FBS{ Few--Body Systems\,\,}
\def\PLB{{ Phys. Lett.\,\,} \bf B}

\def\EPJA{{ Eur. Phys. J.\,\,}\bf A}

\def\PR{ Phys. Rep.\,\,}

\def\IJMPE{{ Int. J. Mod. Phys.\,\,} \bf E}

\def\HI{ Hyperfine Interactions\,\,}
\def\SJPN{ Sov. J. Part. Nucl.\,\,}
\def\APNY{ Ann. Phys. (N.Y.)\,\,}

\def\NC{ Nuovo Cim.\,\,}
\def\RMP{ Rev. Mod. Phys.\,\,}
\def\ANP{ Adv. Nucl. Phys.\,\,}
\def\JPG{ J. Phys.\,\, \bf G}
\def\AJ{ Astrophys. J.\,\,}
\def\CJPB{{ Czech. J. Phys.\,\,} \bf B}
\def\PPNP{Prog. Part. Nucl. Phys.\,\,}

\def\P{Physics\,\,}

\begin{document}
\draft
\title{
{\large{\bf Astrophysical Neutrino Reactions and Muon Capture in
Deuterium}}\footnote{Extended version of the talk, presented
24.11.2010 by E.\,T.\, at the Colloquium, PSI, Villigen,
Switzerland}}
\author{P. Ricci}
\affiliation{Istituto Nazionale di Fisica Nucleare, Sezione di
Firenze, I-50019, Sesto Fiorentino (Firenze), Italy }
\author{E.~Truhl\'{\i}k}
\affiliation{Institute of Nuclear Physics ASCR, CZ--250 68
\v{R}e\v{z}, Czech Republic }

\begin{abstract}

We discuss the importance of precise study of muon capture in
deuterium for correct understanding of some fundamental
astrophysical processes.

\end{abstract}

\noi 

\noi \hskip 1.9cm 

\maketitle

\section{Why the Sun shines}
\label{intro}

The weak nuclear interaction plays crucial role in the formation of
stars in our Universe: it starts the pp chain in stars of the size
of our Sun \cite{AH,HB}, which is the source of its energy. This
chain  has three branches, ppI, ppII and ppIII \cite{AEA}. The ppI
chain is \bea
p\,+\,p\,&\rightarrow&\,d\,+\,e^+\,+\,\nu_e\,,  \label{ppI} \\
p\,+\,p\,+\,e^{-}\,&\rightarrow&\,d\,+\,\nu_e\,,  \label{pepI}\\
d\,+\,p\,&\rightarrow&\,^{3}He\,+\,\gamma\,,    \label{dpf}\\
^{3}He\,+\,^{3}He\,&\rightarrow&\,^{4}He\,+\,2p\,. \label{3he3he}
\eea
In its turn, the chain ppII is
\bea
^{3}He\,+\,^{4}He\,&\rightarrow&\,^{7}Be\,+\,\gamma\,,  \label{3he4he}\\
^{7}Be\,+\,e^{-}\,&\rightarrow&\,^{7}Li\,+\,\nu_e\,,  \label{7beeII}\\
^{7}Li\,+\,p\,&\rightarrow&\,2\,^{4}He\,,   \label{7lip}
\eea
whereas the ppIII chain is
\bea
^{7}Be\,+\,p\,&\rightarrow&\,^{8}B\,+\,\gamma\,,  \label{7bep}\\
^{8}B&\rightarrow&\,^{8}Be^*\,+\,e^+\,+\,\nu_e\,,  \label{8bIII} \\
^{8}Be&\rightarrow&\,2\,^{4}He\,.   \label{8be} \eea Besides, the so
called hep reaction takes place \be
p\,+\,^{3}He\,\rightarrow\,^{4}He\,+\,e^+\,+\,\nu_e\,.  \label{hepI}
\ee In these chains, the following reactions occur, triggered by the
weak nuclear interaction, \bea
p\,+\,p\,&\rightarrow&\,d\,+\,e^+\,+\,\nu_e\,,  \label{pp} \\
p\,+\,p\,+\,e^{-}\,&\rightarrow&\,d\,+\,\nu_e\,,  \label{pep}\\
p\,+\,^{3}He\,&\rightarrow&\,^{4}He\,+\,e^+\,+\,\nu_e\,,  \label{hep}\\
^{7}Be\,+\,e^-\,&\rightarrow&\,^{7}Li\,+\,\nu_e\,,   \label{7bee} \\
^{8}B\,&\rightarrow&\,^{8}Be^*\,+\,e^+\,+\,\nu_e\,.   \label{8b}
\eea The total neutrino flux from the Sun at the surface of Earth is
$\approx$ 6.4 $\times$ 10$^{10}$/cm$^2$ s. The neutrinos produced in
the reaction (\ref{8b}) have a continuous spectrum with the maximum
energy 15 MeV. They have recently been registered in the SNO
detector \cite{SNOI,SNOII,SNOIII} via the reactions \bea
\nu_x\,+\,d\,&\rightarrow&\,\nu^\prime_x\,+\,n\,+\,p\,,  \label{NC} \\
\nu_e\,+\,d\,&\rightarrow&\,e^-\,+\,p\,+\,p\,,  \label{CC} \eea
induced by the weak nuclear interaction, too.  The measured total
flux of active-flavor neutrinos is $\approx$ 5 $\times$
10$^6$/cm$^2$s, whereas the flux of electron neutrinos is $\approx$
1.7 $\times$ 10$^6$/cm$^2$s. The neutral current to charged current
ratio \cite{SNOII} established unambiguously the presence of an
active neutrino flavor other than $\nu_e$ in the observed solar
neutrino flux, thus confirming definitely the phenomenon of the
neutrino oscillations and that the neutrinos possess a finite mass.
When the data from all solar neutrino experiments is combined with
the KamLAND data \cite{KL}, one obtains  \mbox{$\theta_{12}$=34.4$^{+1.3}_{-1.2}$
degrees} and \mbox{$\Delta$m$^2_{12}$=7.59$^{+0.19}_{-0.21} \times$ 10$^{-5}$ eV$^2$} \cite{RM}.

\section{Weak reactions in laboratory}
\label{lab}

However, the reactions (\ref{pp})-(\ref{CC}) cannot be studied
experimentally with the desired accuracy in terrestrial conditions
at present. In order to grasp them, one should address other weak
processes in few-nucleon systems that are feasible in laboratories,
such as \bea
n\,&\rightarrow&\,p\,+\,e^-\,+\,\nu_e\,,     \label{n} \\
\mu^-\,+p\,&\rightarrow&\,n\,+\,\nu_\mu\,, \label{mup} \\
\mu^-\,+\,d\,&\rightarrow&\,n\,+\,n\,+\nu_\mu\,.   \label{mudc} \\
^{3}H\,&\rightarrow&\,^{3}He\,+\,e^-\,+\,\bar{\nu}_e\,,   \label{3h} \\
\mu^-\,+\,^{3}He\,&\rightarrow&\,^{3}H\,+\,\nu_\mu\,,  \label{mu3hec}
 \eea
The one-nucleon weak reactions (\ref{n}) and (\ref{mup}) are now
experimentally and theoretically well explored. The neutron lifetime
is $<\tau>_{world\, av.}$ = 879.9 $\pm$ 0.9 s \cite{SF} and the
singlet capture rate for the reaction (\ref{mup}), $\Lambda_s$ =
725.0 $\pm$ 17.4 s$^{-1}$, has been measured by the MuCap
Collaboration at PSI \cite{MuCap}. As to reactions (\ref{3h}) and
$^{3}$He($\mu^-,\nu_\mu$)$^{3}$H (\ref{mu3hec}), they have been
studied in great detail as well. As a result, the half-life of the
triton is known with an accuracy $\sim$ 0.3 \%, $(fT_{1/2})_t=
(1129.6\pm 3)$ s \cite{AkM}, and the capture rate of the reaction
$^{3}$He($\mu^-,\nu_\mu$)$^{3}$H, $\Lambda_0$= 1496 $\pm$ 4 s$^{-1}$
\cite{AAV,PA} is also known with the same accuracy. The situation
with the reaction $^{2}$H($\mu^-,\nu_\mu$)nn (\ref{mudc}) is less
favorable so far. Indeed, the last measurements of the doublet
capture rate provided $\Lambda_{1/2}$= 470 $\pm$ 29 s$^{-1}$
\cite{M} and $\Lambda_{1/2}$= 409 $\pm$ 40 s$^{-1}$ \cite{CEAL}. As
we shall discuss in this talk, the experiment planned by MuSun
Collaboration \cite{MSE}, which intends to measure $\Lambda_{1/2}$
with an accuracy of \mbox{$\sim$ 1.5 \%,} will help to clarify
essentially the situation in the theory of reactions triggered by
weak nuclear interaction in few-nucleon systems.

\section{Nuclear matrix element of the weak interaction}
\label{nme}

In order to describe these semi-leptonic reactions one should
know how to calculate the matrix element of the weak Hamiltonian
H$_W$ between the initial and final nuclear states, $|i>$ and
$|f>$, respectively. Such a matrix element enters cross sections
and capture rates. One can write generally, \be <f|\,{\hat
H}_W\,|i>\,=\,-\frac{G_W}{\sqrt 2}\,\int \,d\vec x e^{i\vec q
\cdot\vec x}\, [ \vec j(0) \cdot \vec J^a_W(\vec
x)_{fi}\,-\,j_0(0) J^a_{W\,,0}(\vec x)_{fi} ]. \label{MEHW} \ee
The weak lepton current $j_\mu$ is the four-vector, given in the
V-A theory of the weak interactions, e.g., for the muon capture
\be j_\mu(0)\,=\,i {\bar u}(\vec \nu)\,\gamma_\mu\,(1+\gamma_5)\,
 u(\vec k)\,. \label{wlc}
\ee The Dirac spinor $u(\vec k)$ corresponds to the initial muon
with the momentum $\vec k$ and the spinor ${\bar u}(\vec \nu)$
corresponds to the final muon neutrino with the momentum $\vec
\nu$.

Looking at the matrix element (\ref{MEHW}) we conclude that the
problem lies in constructing the hadron currents and the nuclear
wave functions. In its turn, the isovector one-nucleon weak
current $J^a_{W,\,\mu}$ is also of the V-A form, \be
J^a_{W,\,\mu}(q_1)\,=\,J^a_{V,\,\mu}(q_1)\,+\,J^a_{A,\,\mu}(q_1)\,,
\label{JW} \ee where the vector part is given by the matrix
element
 of the isovector
Lorentz 4--vector current operator between the nucleon states,
\be
\hat{J}^a_{V,\,\mu}(q_1) = i \left(g_V(q^{\,2}_1)\gamma_\mu - \frac{g_M(q^{\,2}_1)}{2M}
\sigma_{\mu\,\nu}
q_{1\,\nu} \right)\, \frac{\tau^a}{2}\,,   \label{Vamu}
\ee
and the axial--vector part is analogously,
\be
\hat{J}^a_{A,\,\mu}(q_1) = i \left(-g_A(q^{\,2}_1)\gamma_\mu \gamma_5
+ i \frac{g_P(q^{\,2}_1)}{m_l}
 q_{1\,\mu}
\gamma_5 \right)\,\frac{\tau^a}{2}\,.  \label{Aamu} \ee Here $a$
is the isospin index,  $M$ ($m_l$) is the nucleon (lepton) mass
and the 4--momentum transfer is given by $q_{1\,\mu} = p'_\mu -
p_\mu$, where $p'_\mu$ ($p_\mu$) is the 4--momentum of the final
(initial) nucleon.

Theoretically, all form factors entering the weak one-nucleon
current were well understood and experimentally settled but one
\cite{AM,KA,LI}. It was the induced pseudoscalar g$_P$ that
resisted for quite a long time. Only recently, its value has been
fixed in the already mentioned MuCap experiment,
g$^{exp}_P(q^2$\,=\,0.88\,m$^2_\mu)$ = 7.3 $\pm$ 1.1 \cite{MuCap,CMS},
which is in excellent agreement with the PCAC and chiral perturbation
theory ($\chi$PT) prediction,
\mbox{g$^{th}_P(q^2$\,=\,0.88\,m$^2_\mu)$ = 8.2 $\pm$ 0.2 \cite{CMS,ADo}}.

\section{Chiral symmetry, soft pions}
\label{chs}

The essence of the problem with the study of the above quoted
nuclear reactions is that the quantum chromodynamics (QCD) is
non-perturbative at low energies. The way of handling this obstacle
has already been outlined 50 years ago by realizing that the
description of the electro-weak interaction with a nuclear system,
containing nucleons and pions, should be based on the spontaneously
broken global chiral symmetry SU(2)$_L \times$ SU(2)$_R$
\cite{AD,AFFR}, reflected in the QCD Lagrangian \cite{DGH,WQFTII}.
This fundamental concept was realized after studying the commutation
relations of the charges corresponding to the lepton currents for
the system of electrons and muons with zero masses. In this case,
the lepton charges satisfy the commutation relations of the group
SU(2) $\times$ SU(2) \cite{JB}.

The lepton charges are defined as
\bea
\vec Q_l\,&=&\,\int \Psi^+\,\frac{\vec \tau}{2}\,\Psi d^3 r\,, \label{VC} \\
\vec Q_{5l}\,&=&\,\int \Psi^+\,\gamma_5\,\frac{\vec \tau}{2}\,\Psi
d^3 r\,, \label{AC} \eea
where $\vec \tau$ is the lepton isospin.

From Eqs.\,(\ref{VC}) and (\ref{AC}) one obtains
\be
[\,Q_{il}\,,\,Q_{jl}\,]\,=\,i\epsilon_{ijk}Q_{kl}\,,  \label{CR1}
\ee \be [\,Q_{5il}\,,\,Q_{5jl}\,]\,=\, i\epsilon_{ijk}Q_{kl}\,,
\label{CR2} \ee \be [\,Q_{il}\,,\,Q_{5jl}\,]\,=\, i \epsilon_{ijk}
Q_{5kl}\,.  \label{CR3}
\ee

Gell-Mann made an assumption \cite{MGM} that the universality of the
weak interactions for the leptons and hadrons declares itself in
that the charges connected with the hadron currents satisfy the same
simultaneous commutation relations as the lepton currents.

Besides, the vector part of the weak nucleon current should
satisfy the Conserved Vector Current (CVC) hypothesis,
that permits the identification of the weak vector current with
the isovector part of the electromagnetic current. In its turn the
weak axial nucleon current should satisfy the Partially
Conserved Axial Current (PCAC) hypothesis, \be q_{1\,,\mu}
\hat{J}^a_{A,\,\mu}(q_1) = if_\pi\, m^2_\pi \,\Delta
^\pi_F(q^2_1)\,M^a_\pi\,. \label{PCAC} \ee Here f$_\pi$ is the
pion decay constant, $m_\pi$ is the pion mass, $\Delta
^\pi_F(q^2_1)$ = 1$/(m^2_\pi+q^2_1)$ is the pion propagator and
$M^a_\pi$ is the matrix element of the pion production/absorption
amplitude between the one-nucleon states. It is seen that
the axial current is conserved in the limit of zero
pion mass.

This development induced a burst of calculations and powerful
low-energy theorems for the weak- and electro-production of pions on
nucleon at the threshold were established \cite{AD,AFFR}. Since this
concept is correct only for pions with $q_1$ = 0, the pions are
called soft. Let us note that the results based on current algebras
are model independent.

The spontaneously broken global chiral symmetry SU(2)$_L \times$
SU(2)$_R$ predicts the existence of three massless particles with
the quantum numbers 0$^-$ in the ground state \cite{JG}. Since this
symmetry is broken by the finite mass of the quarks, also these
particles acquire finite mass, and they can be identified with
pions.

Let us note that Eq.(\ref{MEHW}) is valid also for the nuclear
processes with the electromagnetic lepton and nuclear currents.

\section{Impulse Approximation}

Applying Eq.\,(\ref{MEHW}) in calculations of the processes in
nuclei, one first supposed that the nuclear current is approximated
by the sum of the one-nucleon currents (\ref{JW}). This
approximation is called the Impulse Approximation (IA). Generally,
this concept worked well at low energies but it was found that, in
some cases, it failed to describe the data. First it happened  in
the case of thermal neutron capture by proton, \be
n\,+\,p\,\rightarrow\,d\,+\,\gamma\,,  \label{npdg} \ee that the
precise experimental value of the cross section, $\sigma^{exp}$ =
334.2 $\pm$ 0.5 mb\footnote{1 barn (b) = 10$^{-24}$ cm$^2$}
\cite{CWC}, is larger by $\approx$ 10 \% than the IA cross section,
$\sigma^{IA}$ = 302.5 $\pm$ 4.0 mb \cite{HPN}.

This reaction is triggered by the space component of the
electromagnetic isovector current, which is of the order $\sim\,\cal
O$(1/M).

\section{Meson Exchange Currents}
\label{MECs}

Here for the first time, meson exchange currents (MECs) rescued the
situation. The pion production amplitude, evaluated in the soft pion
limit, provided MECs (see Fig.\,1c and Fig.\,1d) that removed about
70 \% of the discrepancy \cite{CR,RB,JFM,DOR,EW}.

\begin{figure}[h!]
\centerline{
\epsfig{file=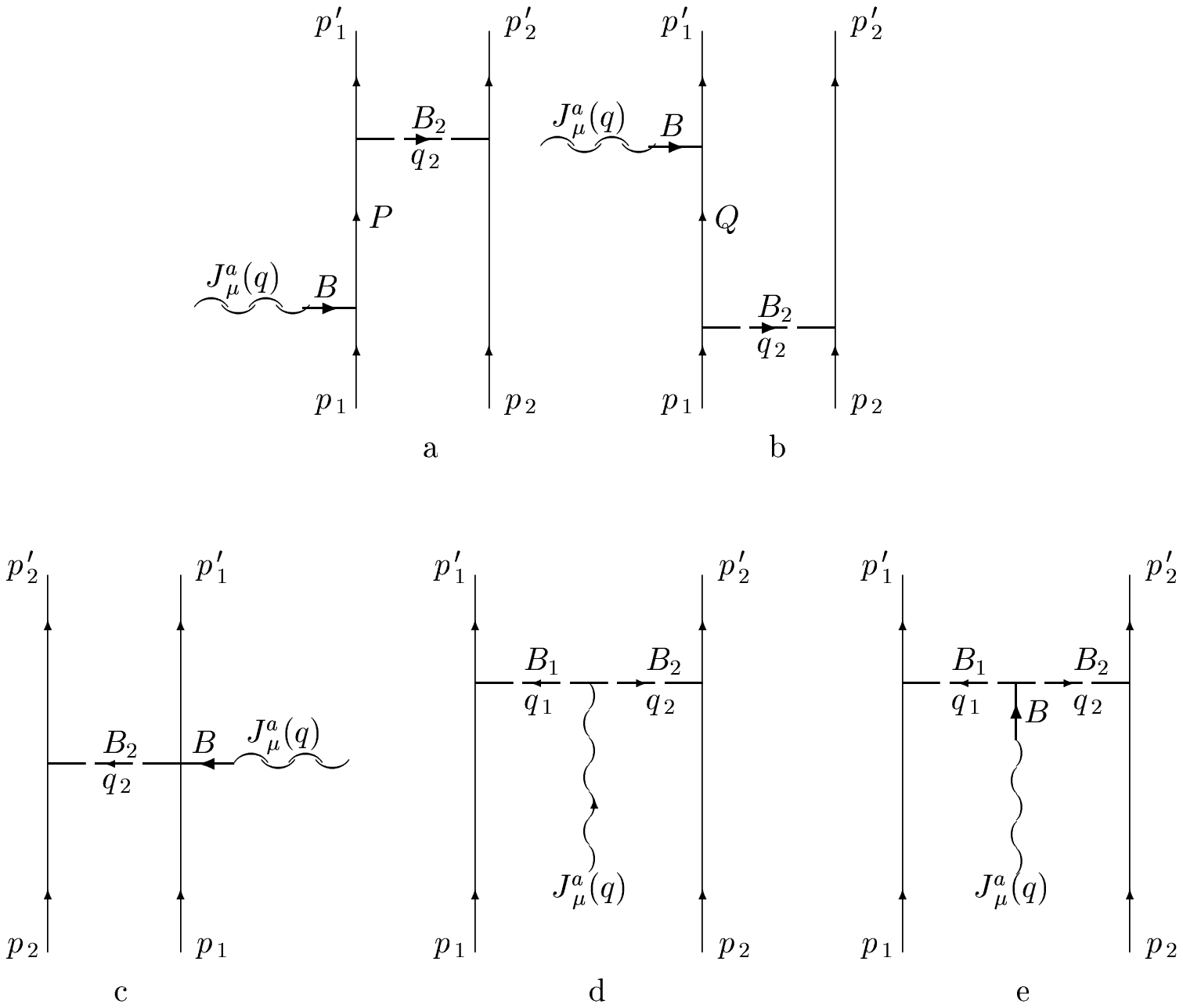}
}
\vskip 0.4cm
\caption{
The possible structure of the two--nucleon current operators;
}

\label{figg1}
\end{figure}

Possible presence of the two-nucleon currents follows also  from
the potential description of the two-nucleon system that we
consider for the sake of simplicity. In this case, the nuclear
Hamiltonian is \be H\,=\,T\,+\,V\,,  \label{H} \ee where T is the
kinetic energy and V is the nuclear potential. For the
electromagnetic current, $J_\mu(q)$, the current conservation
reads, \be \vec q \cdot \vec J(\vec q)\,=\,[\,H\,,\,\rho(\vec
q)\,]\,.   \label{cc} \ee Writing the current as the sum of the
one-nucleon and two-nucleon parts, \be J_\mu(q)\,=\,\sum^2_1\,
J_\mu(1,i,q_i)\,+\,J_\mu(2,q)\,,   \label{cur} \ee one gets \bea
\vec q_i \cdot \vec J(1,i,\vec q_i)\,&=&\,[\,T_i\,,\,\rho(1,i,\vec q_i)\,]\,,\quad i=1,\,2\,, \label{obc}\\
\vec q \cdot \vec J(2,\vec q)\,&=&\,[\,T_1+T_2\,,\,\rho(2,\vec q)\,]\,+\,([\,V\,,\,\rho(1,1,\vec q_1)\,]
\,+\,(1\leftrightarrow 2))\,,   \label{tbc}
\eea
where $\vec q$ = $\vec q_1$ + $\vec q_2$.

It is seen from Eq.\,(\ref{tbc}) that this equation cannot be
fulfilled if the MECs are absent.

\section{Chiral Lagrangians, hard pions}
\label{chl}

As we have already mentioned, the approach of soft pions is valid at
the threshold energies. Next we discuss the concept of chiral
Lagrangians allowing to go beyond this restriction. The approach of
chiral Lagrangians is based on an assumption that any Lagrange
theory satisfying the requirements \\
1.\, The constructed currents satisfy the fundamental commutation
relations
known from the soft-pion approach. \\
2.\, In the limit of zero pion mass the weak axial current is exactly conserved, \\
should reproduce the results of the soft-pion technique \cite{AFFR}.
This is possible thanks to the fact that the correct results are
obtained already at the level of trees (no loops). The chiral
Lagrangians of the pion-nucleon system reflect the global chiral
symmetry. Standardly, they are constructed in non-linear realization
of the chiral symmetry \cite{WNLL}.

One can also consider the Lagrangians reflecting the local chiral
symmetry \cite{JSC,OZ}. This step allowed to extend the nuclear
system by the $\rho$- and a$_1$ mesons, as compensating Yang-Mills
fields. However, the problem was that the heavy meson masses
violated the symmetry. Later on, a concept of hidden local symmetry
allowed one to avoid this conceptual difficulty \cite{HLSLM,BKY}.
The nonlinear hard pion chiral Lagrangians \cite{OZ,ITI,STG} served
then as starting point for constructing the one-boson exchange
currents in the tree approximation. We shall call this concept as
the Tree Approximation Approach (TAA).

This approach was also applied to construct the one-boson exchange
potentials \cite{BS,MHE,MANUP,SKTS,CDB}. Besides the potentials of
this sort, many phenomenological potentials of various quality were
also constructed \cite{HJ,RPOT,SSC,PAR,AV18}. The precise second
generation potentials Nijmegen I \cite{SKTS}, CD-Bonn \cite{CDB} and
AV18 \cite{AV18} have $\chi^2\,\sim$ 1.

Let us note that in order to make realistic calculations, both the
one-boson MECs and potentials should be supplied with the strong
form factors by hands. Fortunately, this can be done in such a way
that the CVC and PCAC hypotheses are still valid. Consequently, one
can describe the MECs effect consistently: both the MECs and nuclear
potentials are obtained within the same approach.

The reactions (\ref{NC})-(\ref{CC}) and (\ref{mudc})-(\ref{mu3hec})
were studied within TAA in
Refs.\,\cite{NSAPMGK,MRTI,MRTV,CiT,CT,ATCS,TKK,RTMS}. These
reactions are triggered by the space component of the weak current.
Its one-nucleon weak axial part is of the order $\sim\,\cal O$(1),
whereas the space component of the weak axial MECs is $\sim\,\cal
O$(1/M$^2$). This fact makes the calculations of the weak axial MECs
effects difficult.

Here we present for the reaction  $^{2}$H($\mu^-,\nu_\mu$)nn the
results, calculated with the first generation potentials, for the
doublet capture rate $\Lambda_{1/2}$ = 416 $\pm$ 7 s$^{-1}$
\cite{ATCS} and $\Lambda_{1/2}$ = 397.8 - 399.6 s$^{-1}$ \cite{TKK}.
The total effect of the weak MECs was estimated as $\sim$ 30.6 -
33.1 s$^{-1}$ \cite{TKK}. Recent calculations with the precise
potential Nijmegen I provided $\Lambda_{1/2}$ = 416 $\pm$ 6 s$^{-1}$
\cite{RTMS}. Analogous calculations for the reaction
$^{3}$He($\mu^-,\nu_\mu$)$^{3}$H provided for the statistical rate
$\Lambda_0$ = 1502 $\pm$ 32 s$^{-1}$ \cite{CT} and $\Lambda_0$ =
1484 $\pm$ 8 s$^{-1}$ in a purely phenomenological approach
\cite{MSRKV}.

\begin{figure}[h!]
\centerline{
\epsfig{file=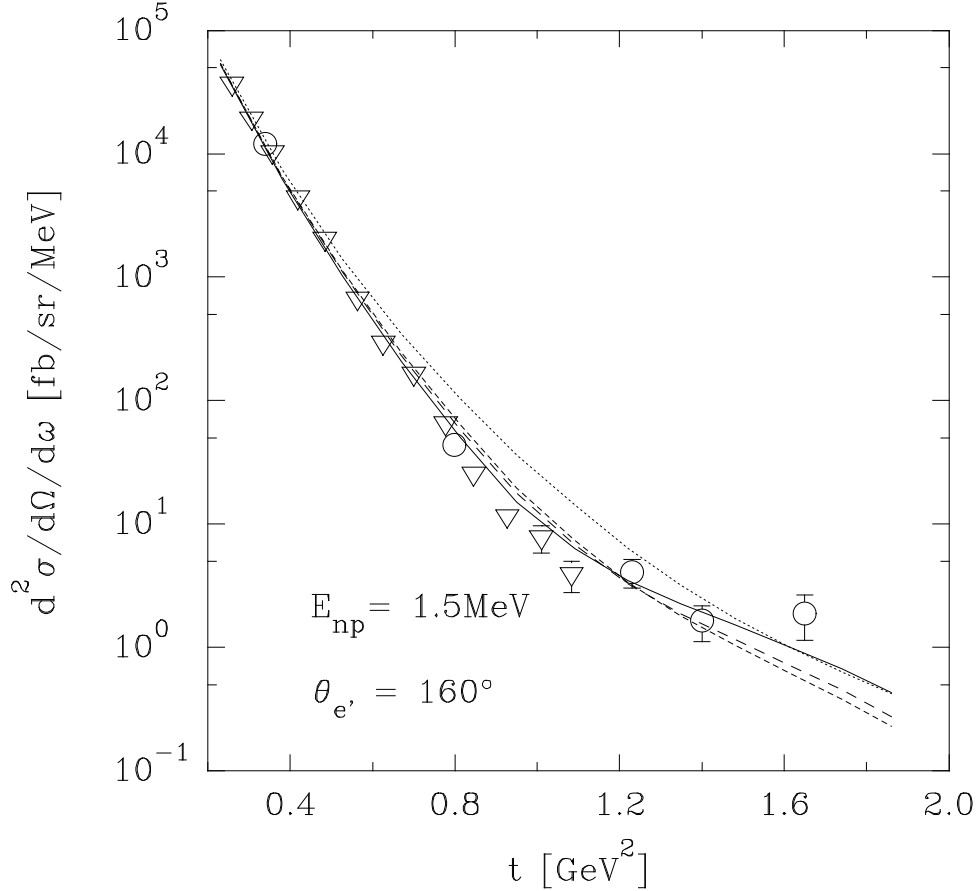}
}
\vskip 0.4cm
\caption{
The double differential cross section for the process (\ref{bded}).
}

\label{figg2}
\end{figure}

Generally, within this approach one can describe  well the nuclear
phenomena, triggered by the electro-weak interaction, up to energies
$\sim$ 1 GeV \cite{ITR,KTR,ITPR,JFM,DOR,EW,CSc,DFM,SCK,GIG}. In
Fig.\,\ref{figg2}, we present the double differential cross section
for the reaction of the backward deuteron
\mbox{electro-disintegration \cite{STG},} \be
e\,+\,d\,\rightarrow\,e'\,+\,n\,+\,p\,, \label{bded} \ee where the
energetic electrons are scattered backwards, closely to 180 degrees.
The detailed analysis of the contributions of particular mesons to
the cross section has been made in Refs.\,\cite{TA,TS}.

From 1980's, in parallel with the above discussed schemes of the
calculations in nuclear physics, new approach was worked out,
providing more general framework for systematic construction of many
body currents and potentials. The price for this was explicit
absence of all degrees of freedom in the Lagrangian but nucleons and
pions and applicability to nuclear phenomena only in the low energy
region.

\section{Effective field theory}
\label{eft}

The fundamental step allowing to go beyond the tree approximation
was made by Weinberg in 1979 \cite{W}. In this work, Weinberg
formulated an effective field theory (EFT): if one writes down the
most general possible Lagrangian, including all terms consistent
with assumed symmetry principles, and then calculates matrix
elements with this Lagrangian to any given order of perturbation
theory, the result will be the most general possible S-matrix
consistent with perturbative unitarity, analyticity, cluster
decomposition, and the assumed symmetry properties.

So in this way, one can construct dynamical theory, not limited to
the tree approximation.

At low energies, the effective degrees of freedom in nuclear physics
are pions and nucleons, rather than quarks and gluons. With the
heavy mesons and nucleon resonances integrated out and the
spontaneously broken global chiral symmetry as assumed symmetry, one
obtains the $\chi$PT of the nucleon-pion system. The resulting
effective Lagrangian is given by a string of terms, dictated by the
chiral symmetry, with increasing chiral dimension \cite{FMMS}, \be
{\cal L}_{\pi N}\,=\,{\cal L}^{(1)}_{\pi N}\,+\,{\cal L}^{(2)}_{\pi
N}\,+\,{\cal L}^{(3)}_{\pi N}\,+\,{\cal L}^{(4)}_{\pi N}\, +\,...\,.
\label{CHPTL} \ee This is the QCD Lagrangian of the $\pi$N system at
low energies. The Lagrangian of dimension one is simple \be {\cal
L}^{(1)}_{\pi N}\,=\,\bar \Psi ( D_\mu
\gamma_\mu\,-\,M\,+\,i\frac{g_A}{2}\,u_\mu
\gamma_\mu\,\gamma_5)\Psi\,, \label{DOL} \ee where $D_\mu$ and
$u_\mu$ depend non-linearly on the pion field and external
interactions, and $\gamma$s are the Dirac matrices. Besides, $g_A$
is the weak interaction constant.

At the second order, seven independent terms appear, at the third
order, one has 23 independent terms, and at the dimension four,
there are 118 independent operators \cite{FMMS}. Each term is
multiplied by a (low energy) constant (LEC) that is fixed either in
the process of the elimination of heavier resonances, or by the
data.

Weinberg also established counting rules \cite{W1,W2} allowing one
to classify the importance of contribution of various diagrams into
perturbative expansion of the S-matrix in positive powers $\nu$ of
$Q/\Lambda_\chi$, where $Q$ is a quantity characterizing the hadron
system (momentum, energy or the pion mass) which is small in
comparison with the heavy scale $\Lambda_\chi\,\sim$ 1 GeV.

As we have already noted above, in order to calculate reliably the
capture rates and cross sections of reactions one needs to know
accurately the nuclear wave functions (potentials) and the current
operators.

\subsection{Space component of the weak axial MECs}
\label{scwam}

\begin{figure}[h!]
\centerline{
\epsfig{file=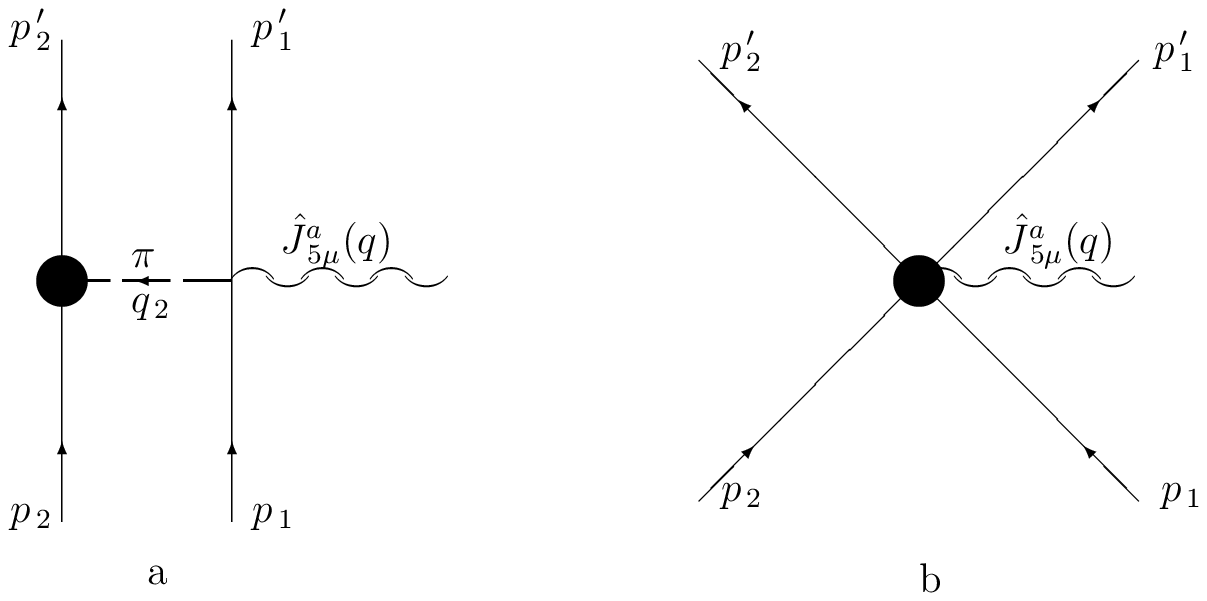}
}
\vskip 0.4cm
\caption{
The general structure of the two--nucleon weak axial operators;
a-- the long-range operator, b-- the short-range operator.
}

\label{figg3}
\end{figure}

The weak axial currents were constructed within the $\chi$PT in
Refs.\,\cite{PKMR,DG}. At the leading ($\nu$=0)- (LO) and the
next-to-leading ($\nu$=1) (NLO) orders, the weak nuclear current
consists of the well-known single nucleon terms. The space component
of the weak axial MECs, which is of the main interest here because
it enters in calculations of observables in all weak reactions
mentioned above, appears at N$^{3}$LO ($\nu=3$). We already know
that the single nucleon current is known precisely. On the other
hand, in the space component of the weak axial MECs appears one LEC,
called $\hat d^R$. This parameter manifests itself in the
two-nucleon contact vertex with the weak axial current (see
Fig.\,\ref{figg3}). Besides, it is present also in a contact term
part of the three-nucleon force constructed within the $\chi$PT. It
follows that if this LEC will be fixed in one of the few-nucleon
processes feasible in the laboratory, one can make model-independent
predictions for other weak processes triggered by this component of
the weak axial current.

\subsection{The hybrid calculations}
\label{hc}

Up to present only few calculations fulfil the requirement of
consistency and so called 'hybrid' approach is applied instead: the
current operator is constructed within the $\chi$PT as outlined
above, but the wave functions are generated either from the
one-boson-exchange- or phenomenological potentials of the TAA
approach. Besides, in calculating the observables for the
three-nucleon processes, also the three-nucleon forces
Tucson-Melbourn\cite{CEA} and Urbana IX \cite{PEA} were used.

So far almost all calculations aiming to study the weak interaction
in few-nucleon systems profited from the precise knowledge of the
half-life of the triton to extract the LEC $\hat d^R$. In this way,
in Ref.\,\cite{ASP}, this constant  was extracted from the
Gamow-Teller (GT) matrix element for the reaction (\ref{3h}), and
then the spectroscopic factors\footnote{The spectroscopic factor is
defined as $ S(E)\,=\,\sigma(E)\,E\,e^{2\pi \eta}$, where $E$ is the
energy, $\sigma$ is the cross section, and $\eta$ is the Sommerfeld
parameter describing the barrier penetrability. }
S$_{pp}$(0)=3.94$\times$(1$\pm$0.004)$\times$10$^{-25}$ \mbox{MeV b}
and S$_{hep}$(0)=(8.6$\pm$1.3)$\times$10$^{-30}$ keV b were
calculated for reactions of the proton-proton fusion (\ref{pp}) and
the hep reaction (\ref{hep}), respectively.

Using the same values of $\hat d^R$, in Ref.\,\cite{APKM} the
doublet capture rate, $\Lambda_{1/2}$=386 s$^{-1}$, for the reaction
$^{2}$H($\mu^-,\nu_\mu$)nn was obtained and in Ref.\,\cite{ASPFK}
the cross sections of the $\nu$d reactions, (\ref{NC}) and
(\ref{CC}), were calculated. Similarly, in Ref.\cite{GAZ}, the
capture rate $\Lambda_0$= 1499$\pm$16 s$^{-1}$ for the reaction
$^{3}$He($\mu^-,\nu_\mu$)$^{3}$H was gained. In Ref.\cite{PI1},
Marcucci and Piarulli used AV18 NN potential to generate the nuclear
wave functions for the process $^{2}$H($\mu^-,\nu_\mu$)nn and AV18 +
Urbana IX NNN potentials for the reaction
$^{3}$He($\mu^-,\nu_\mu$)$^{3}$H. The weak current was taken from
the $\chi$PT approach with the potential current of Ref.\,\cite{RMT}
added. The calculations  resulted in $\Lambda_{1/2}$ = 393.1 $\pm$ 8
s$^{-1}$ and $\Lambda_0$ = 1488 $\pm$ 9 s$^{-1}$.

The problem with these calculations is that in the three-nucleon
systems, the consistent calculations require to know two LECs, c$_D$
and c$_E$ \cite{GQN}. The constant  c$_D$ is related to the constant
$\hat d^R$ as \be \hat d^R\,=\,\frac{M_N}{\Lambda_\chi
g_A}c_D\,+\,\frac{1}{3}M_N\,(\hat c_3\, +\,2\hat
c_4)\,+\,\frac{1}{6}\,.  \label{reldrcd} \ee Here the LECs $\hat
c_3$ and $\hat c_4$, together with $\hat c_1$, $\hat c_2$ and c$_6$,
are obtained either from the $\pi$N- \cite{VB} or NN scattering
\cite{EM}, or in the process of elimination of higher resonances
from the general $\chi$PT Lagrangian \cite{BKM1}.

In the hybrid calculations of the process
$^{3}$He($\mu^-,\nu_\mu$)$^{3}$H mentioned above, it is not possible
to fix c$_D$ and c$_E$ simultaneously, because c$_E$ does not enter
them. Besides, the constant c$_D$ ($\hat d^R$) enters not only the
weak axial MECs, but also the contact and one-pion exchange part of
the three-nucleon force constructed within the $\chi$PT, whereas the
constant c$_E$ enters the three-nucleon contact term \cite{UVC,EEA}.

Let us note also the work \cite{RTMS}, where  the capture rate
$\Lambda_{1/2}$ = 416 $\pm$ 6 s$^{-1}$ for the reaction
$^{2}$H($\mu^-,\nu_\mu$)nn was calculated within the TAA in
various current models and from this rate, the model dependence of the
value of $\hat d^R$ in hybrid calculations was studied.

\subsection{Chiral potentials}
\label{chp}

The early theory of nuclear force also started from the nucleon-pion
system in 1950s but failed to describe reasonably well the empirical
data. As we have already discussed,  only later the chiral symmetry
was recognized as proper symmetry of the Nature. In 1960s, the heavy
mesons were discovered and used to construct quite successful models
of nuclear force. The price for it was the necessity to introduce
phenomenological strong form factors into the one-boson-exchange
potentials and to deal with a not well known scalar-isoscalar
$\sigma$ meson.

With the advent of QCD, the attempts to construct the nuclear
forces from the QCD-inspired quark models appeared. But only
within the $\chi$PT it was possible to derive consistently the
nuclear force between the N nucleons from the same Lagrangian and
precise two-nucleon potentials have been constructed up to
N$^{3}$LO: the Entem-Machleidt (EM) \cite{EM} and
Epelbaum-Gloeckle-Mei\ss ner \cite{EGM} potentials. In both cases,
the entering parameters are standardly extracted from the fit to
the nucleon-nucleon scattering data and the deuteron properties,
or some of them are adopted from the analysis of the $\pi$N
scattering \cite{EGM}.

The three-nucleon force has been derived within the $\chi$PT, but
only up to N$^2$LO \cite{UVC,EEA} so far. As we already know, this
force contains two constants, c$_D$ and c$_E$, to be determined. The
first consistent extraction of the constants c$_D$ and c$_E$ from
the three-nucleon system has recently been made by Gazit, Qualioni
and Navr\'atil in Ref.\,\cite{GQN}, where these constants are
constrained by simultaneous calculations of the binding energies of
the three-nucleon systems and of the triton $\beta$ decay. The
nuclear wave functions are generated in accurate {\it ab initio}
calculations using both the two-nucleon EM \cite{EM} and
three-nucleon N$^2$LO force \cite{UVC,EEA}, whereas the process
(\ref{3h}) is calculated with the weak axial MECs derived from the
same $\chi$PT Lagrangian \cite{GAZ} as the nuclear forces. The
resulting values of the two constants are restricted in the
intervals, -0.3 $\le$ c$_D$ $\le$ -0.1, \mbox{ -0.220 $\le$ c$_E$
$\le$ -0.189}.

The reactions of muon capture in deuterium and in $^{3}$He have very
recently been calculated by the Pisa group (MEAL) \cite{PI2} in the
TA- and hybrid approaches and within the $\chi$PT as well. The
resulting values of the capture rates are, $\Lambda_{1/2}$ = (389.7
- 394.3) s$^{-1}$ and  \mbox{$\Lambda_0$ = (1471 - 1497) s$^{-1}$.}

In order to compare our results with those of MEAL, we choose the
last row of TABLE VI \cite{PI2} in which we divide the items by
1.024 \cite{CMS}. This number represents the inner radiative
corrections taken into account by MEAL. In our analogous $\chi$PT
calculations \cite{RT} with the same N$^3$LO potential \cite{EM}, we
take the value of c$_D$ = -0.2 \cite{GQN} and $\Lambda$ = 500 MeV.
Besides, using the values of the LECs (in unit GeV$^{-1}$), c$_1$ =
-0.81, c$_2$ = 2.80, c$_3$ = -3.20 and c$_4$ = 5.40 \cite{EM}, we
obtain from Eq.\, (\ref{reldrcd}) $\hat d^R$ = 2.400. Let us note
that this value of $\hat d^R$ differs considerably from $\hat d^R$ =
1.00(9), presented in the last row of TABLE V \cite{PI2}.

The results of calculations are given in Table \ref{tab:dtr}. As it
is seen, the main difference stems from the $^1$S$_0$ and $^3$P$_1$
waves. Also the contributions for the $^1$D$_2$ wave differ, but
they are small. Since the calculated MECs effect, $\Delta$MEC = 15.4
s$^{-1}$ \cite{RT}, is similar to the one obtained in
\cite{APKM,RTMS}, we conclude that the difference comes from the IA
calculations in the channels $^1$S$_0$ and $^3$P$_1$.
\begin{table}[htb]
\caption{Contributions to $\Lambda_{1/2}$ from partial waves (in
s$^{-1}$). In the last column, all the contributions are summed up.
RT - our calculations, MEAL - the last column of TABLE VI \cite{PI2}
divided by 1.024. }
\begin{center}
\begin{tabular}{|l | c  c  c  c  c  c  c  |}\hline
         & $^1$S$_0$ & $^3$P$_0$ & $^3$P$_1$ & $^3$P$_2$ & $^1$D$_2$ &  $^3$F$_2$ & total \\\hline
RT  & 258.7 & 19.6 & 59.5 & 69.7 & 6.2 & 0.4 & 414.1 \\
MEAL& 244.6 & 19.4 & 45.3 & 69.8 & 4.3 & 0.9 & 384.3\\\hline
\end{tabular}
\end{center}
\label{tab:dtr}
\end{table}

\section{Conclusions}
\label{con}

We have seen that the constant $\hat d^R$ (c$_E$) is currently
extracted from the triton beta decay rate. However, it is abundantly
clear from our discussion that dealing with the complexity of the
three-nucleon system can be avoided, if $\hat d^R$ would be
determined accurately from the muon capture in deuterium. Then one
can consistently calculate other two-nucleon weak processes, such as
proton-proton fusion reaction (\ref{pp}) and both reactions,
(\ref{NC}) and (\ref{CC}), of solar neutrinos with the deuterons, so
important for the astrophysics. Besides it turns out \cite{GF} that
$\hat d^R$ enters also the capture rate for the reaction
\mbox{$\pi^-\,+\,d\,\rightarrow\,\gamma\,+\,2\,n$}, which is the
best source of information on the neutron-neutron scattering length
a$_{nn}$.

\end{document}